\documentclass[submission,copyright,creativecommons]{eptcs}
\providecommand{\event}{iWIGP 2010} 
\usepackage[utf8]{inputenc}
\usepackage{xspace}
\usepackage{color}
\usepackage{amsmath} 
\usepackage{amssymb}
\usepackage{stmaryrd} 
\usepackage{helvet}
\usepackage{courier}
\usepackage{formal}
\usepackage{tikz}
\usetikzlibrary{automata}
\usetikzlibrary{arrows}
\usepackage{color}
\newcommand{\ie}{i.\,e.~}

\usepackage{tabularx}
\usepackage{breakurl}

\begin{document}

\title{A LTL Fragment for GR(1)-Synthesis}

\author{Andreas Morgenstern and Klaus Schneider
\institute{%
	University of Kaiserslautern\\
	P.O. Box 3049 \\
	67653 Kaiserslautern, Germany \\
	email: \{morgenstern,schneider\}@cs.uni-kl.de}
}
\def\titlerunning{A LTL Fragment for GR(1)-Synthesis}

\def\authorrunning{A. Morgenstern and K. Schneider}

\newtheorem{definition}{Definition}
\newtheorem{proposition}{Proposition}
\newtheorem{lemma}{Lemma}
\newtheorem{theorem}{Theorem}
\newtheorem{corollary}{Corollary}
\newtheorem{remark}{Remark}
\newtheorem{example}{Example}
\def\qed{\hfill \rule{2mm}{2mm}}

\maketitle

\begin{abstract}
The idea of automatic synthesis of reactive programs starting from temporal logic (\LTL{}) specifications is quite old, but was commonly thought to be infeasible due to the known double exponential complexity of the problem. However, new ideas have recently renewed the interest in \LTL{} synthesis: One major new contribution in this area is the recent work of Piterman et al. who showed how polynomial time synthesis can be achieved for a large class of \LTL{} specifications that is expressive enough to cover many practical examples. These \LTL{} specifications are equivalent to $\omega$-automata having a so-called GR(1) acceptance condition. This approach has been used to automatically synthesize implementations of real-world applications. To this end, manually written deterministic $\omega$-automata having GR(1) conditions were used instead of the original \LTL{} specifications. However, manually generating deterministic monitors is, of course, a hard and error-prone task. In this paper, we therefore present algorithms to automatically translate specifications of a remarkable large fragment of \LTL{} to deterministic monitors having a GR(1) acceptance condition so that the synthesis algorithms can start with more readable \LTL{} specifications.
\end{abstract}

\section{Introduction}

In the last decades, the influence of computer systems on our everyday life has been constantly growing. As computer systems enter more and more safety-critical areas, their correctness is essentially important to avoid malfunctioning systems. Thus, one of the main challenges in computer science is the design of provably correct systems. Many of these safety-critical computer systems are reactive embedded systems. These are non-terminating systems that interact with their environments during their infinite computations. Typically, concurrency and infinite computations with respect to the environment make it difficult to analyze and design such systems correctly.

There are currently two main approaches to the design of provably correct reactive systems: In the first approach, called \emph{formal verification}, one checks that a manually written implementation satisfies a given specification that is typically formulated in the temporal logic \LTL{} \cite{Pnue77a,Emer90}. In the second approach, called \LTL{} synthesis, a provably correct implementation is automatically derived from the given \LTL{} specification. While formal verification is nowadays even routinely used in safety-critical system designs, \LTL{} synthesis is still immature. Of course, the double exponential complexity of \LTL{} synthesis compared to the single exponential one of \LTL{} model checking is one reason for this situation. We believe, however, that the applicability of tools based on both methods can be significantly improved by better data structures and algorithms.

For example, a major breakthrough in formal verification has been achieved by symbolic representations of states and transitions with propositional formulas which became known as symbolic model checking \cite{BCMD90}. With the advent of these succinct data structures and efficient decision procedures for propositional formulas, it has become possible to verify complex systems. In a similar way, new methods for SAT checking and SMT solvers opened the way to verify even larger systems.

It is natural to try to make use of such data structures and algorithms also for \LTL{} synthesis. However, this is not directly possible, since the currently available \LTL{} synthesis procedures consist of two steps: The first step is the translation of the \LTL{} specification to an equivalent $\omega$-automaton. The usual translation procedures generate a nondeterministic automaton that can be directly used for symbolic model checking. However, nondeterministic automata can, in general, not be used for \LTL{} synthesis. Even though there are pseudo-deterministic automata like the good-for-games automata that can still be used for \LTL{} synthesis, the second step usually consists of a determinization of the obtained automata (since deterministic automata can be definitely used without further restrictions). The problem is, however, that determinization is considerably more complex for $\omega$-automata than for automata on finite words. In particular, a major drawback of the currently known determinization procedures is their explicit representation of the automata that does not make use of symbolic data structures. Since a translation from \LTL{} to deterministic automata may lead to automata having a double exponential size in terms of the length of the formula, explicit state space representations are limited to handle very small \LTL{} formulas.

One possibility to overcome the complexity problem of \LTL{} synthesis is to consider restricted classes of \LTL{}. For example, \cite{AlTo04,Maid00} consider subsets of \LTL{} to obtain deterministic automata with less than double exponential size. Wallmeier et al. \cite{WaHT03} developed a synthesis algorithm to synthesize request-response specifications which are of the form $\ALWAYS{(\varphi_i \rightarrow \EVENTUAL{\psi_i})}$ for multiple $i$ which leads to a synthesis procedure with only exponential complexity. Piterman et.~al proposed in \cite{PiPS06} an approach to synthesize generalized reactivity formulas with rank 1 (abbreviated as GR(1) formulas), \ie formulas of the form ${\left(\bigwedge_{i=0}^N\ALWAYS{\EVENTUAL{\varphi_i}}\right) \rightarrow \left(\bigwedge_{j=0}^M\ALWAYS{\EVENTUAL{\varphi_j}}\right)}$. Their algorithm runs in time $K^3$ where $K$ is the size of the state space of the design. If a collection $\Phi_i$ of \LTL{} formulas representing assumptions on the environment, and a collection $\Psi_j$ of formulas representing conclusions for the system, can all be represented by deterministic Büchi automata, this approach can be used to obtain a synthesis procedure for the entire \LTL{} specification ${\left(\bigwedge_{i=0}^N \Phi_i \right) \rightarrow \left(\bigwedge_{j=0}^M \Psi_j\right)}$. 

The work reported in \cite{PiPS06} has been extensively used. Its feasibility was demonstrated in \cite{BGJP07,BGJP07a,JGWB07} which considers ARM's Advance Micro-System Bus Architecture as well as a case study of a generalized buffer example included in IBM's RuleBase system. In those case studies, an implementation realizing the given formal specification has been derived and has been afterwards converted to a circuit. In fact, those case studies have been the first real-life blocks that have been automatically synthesized from high-level temporal logic specifications. Further applications include usage in the context of production of robot systems \cite{WoTM10}.

The main drawback of previously published works using the GR(1)-approach of Piterman et al. is that the unavoidable determinization step was carried out manually by a human developer, since no tool support for the translation of temporal logic formulas to corresponding $\omega$-automata was available. The translation to deterministic automata is considerably hard in general \cite{KuVa98c} and may introduce errors due to the human intervention.

To eliminate this drawback from the GR(1)-approach, we present in this article a remarkable large subset of \LTL{} that can be translated to sets of deterministic Büchi automata representing the assumptions on the environment and the guarantees a system has to satisfy. To this end, we reconsider the temporal logic hierarchy that has been investigated by Chang, Manna, Pnueli, Schneider and others \cite{MaPn87c,ChMP92,MaPn90,MaPn91,Schn01b,Schn03}. This temporal logic hierarchy defines subsets of \LTL{} that correspond to the well-known automaton hierarchy, consisting of safety, guarantee/liveness, fairness/response/Büchi, persistence/co-Büchi properties as well as their boolean closures (obligation and reactivity properties). Using a syntactic characterization of this hierarchy \cite{Schn01b,Schn03}, we can, in particular, \emph{syntactically} determine for given \LTL{} formulas whether the formula can be represented by a deterministic Büchi automaton. Hence, given a set of formulas representing assumptions and conclusions, we can determine whether they can be used as an input for GR(1)-synthesis. Clearly, since we only check this syntactically, it may be the case that we reject formulas that could be used for GR(1)-synthesis, but we never produce an error. In practice, it turned out that essentially no GR(1) formula is rejected by our syntactic check.

The syntactic approximation to determine GR(1) membership is one contribution of this paper. Another one is the observation that the negation of each formula that can be translated to a deterministic Büchi automaton can be translated to a non-deterministic co-Büchi automaton. It is well-known that non-deterministic co-Büchi automata can be determinized by the Breakpoint construction \cite{MiHa84} that is well-suited for a symbolic implementation \cite{MoSL08,BoKu09a}. From this co-Büchi automaton, we can easily obtain a deterministic Büchi automaton (again via negation, which is trivial for deterministic automata \cite{Schn03}) that is equivalent to the original formula. Hence, our second observation leads to a very efficient translation procedure for the identified \LTL{} formulas to deterministic Büchi and co-Büchi automata.

We have implemented this synthesis procedure that (1) syntactically determines whether a formula can be represented with a GR(1)-property and (2) applies the mentioned symbolic determinization procedure for Büchi/co-Büchi automata. Finally, we apply the GR(1)-synthesis using an existing implementation of the GR(1)-Synthesis approach \cite{BCGH10a}.

\section{Preliminaries}

\subsection{\texorpdfstring{Linear Temporal Logic \LTL{}}{Linear Temporal Logic LTL}}
For a given set of Boolean variables $\Var{}$, we define the set of $\LTL{}$ formulas by the following recursive definition:
\begin{definition}[Syntax of Linear Temporal Logic (\LTL{})]
The set of  LTL formulas over a set of variables $\Var{}$ is the smallest set with the following properties:
\begin{itemize} \setlength{\itemsep}{-1mm}
\item $\True, \False \in \LTL{}$ 
\item $a \in\LTL{}$ for $a \in \Var{}$
\item boolean operators: $\neg \varphi$, $\varphi \wedge \psi$, $\varphi \vee \psi\in \LTL{}$ if $\varphi,\psi \in \LTL{}$
\item future temporal operators: $\NEXT{\varphi}$, $\SUNTIL{\psi}{\varphi}$, $\BEFORE{\psi}{\varphi}$  if $\varphi,\psi \in \LTL{}$
\item past temporal operators: $\PNEXT{\varphi}$, $\PSNEXT{\varphi}$, $\PSUNTIL{\psi}{\varphi}$, $\PBEFORE{\psi}{\varphi}$  if $\varphi,\psi \in \LTL{}$	
\end{itemize}
\end{definition}

\noindent The semantics of \LTL{} can be given with respect to a path through a structure (e.g. an $\omega$-automaton), where a path is an infinite word over the alphabet $2^\Var{}$.

$\NEXT{\varphi}$ holds on a path $\pi$ at position $t_0$ if $\varphi$ holds at position $t_0+1$ on the path. $\SUNTIL{\psi}{\varphi}$ holds at $t_0$ iff $\psi$ holds for some position $\delta\geq t_0$ and $\varphi$ holds invariantly for every position $t$ with $t_0 \leq t < \delta$ \ie $\varphi$ holds \emph{until} $\psi$ holds. The \emph{weak before} operator $\BEFORE{\psi}{\varphi}$ holds at $t_0$ iff either $\varphi$ holds before $\psi$ becomes true for the first time after $t_0$ or $\psi$ never holds after $t_0$. 

In addition to the future time temporal operators, there are also the corresponding past time temporal operators. These are defined analogously with the only difference that the direction of the flow of time is reversed. For example, $\PSUNTIL{\psi}{\varphi}$ holds on a path at position $t_0$ iff there is a point of time $\delta$ with $\delta \leq t$ such that $\psi$ holds on that path at position $\delta$ and $\varphi$ holds for all positions $t$ with $\delta<t \leq t_0$. The past time correspondence of the next-time operator is called the previous operator: $\PSNEXT{\varphi}$ holds on a path at position $t_0$ iff $t_0>0$ and $\varphi$ holds at position $t_0-1$. Additionally, there is a weak variant, where $\PNEXT{\varphi}$ holds on a path at position $t_0$ iff $t_0=0$ holds or $\varphi$ holds at position $t_0-1$.   

Other operators can be defined in terms of the above ones:
\begin{alignat*}{2}
\ALWAYS{\varphi} & =\BEFORE{\neg \varphi}{\False}  & \qquad \PALWAYS{\varphi} & =\PBEFORE{\neg \varphi}{\False} \\
\EVENTUAL{\varphi} & =\SUNTIL{\varphi}{\True}  & \qquad \PEVENTUAL{\varphi} & =\PSUNTIL{\varphi}{\True} \\
\BEFORE{\psi}{\varphi} & =  \neg \SUNTIL{\psi}{\neg \varphi} & \qquad \PBEFORE{\psi}{\varphi} & =  \neg \PSUNTIL{\psi}{\neg \varphi} \\
\UNTIL{\psi}{\varphi} & =  \BEFORE {(\neg \varphi \wedge \neg \psi)}{\psi}  & \qquad \PUNTIL{\psi}{\varphi} & =  \PBEFORE {(\neg \varphi \wedge \neg \psi)}{\psi}\\
\WSBEFORE{\psi}{\varphi} & = \SUNTIL {(\varphi \wedge \neg \psi)}{\neg \psi}   & \qquad \PWSBEFORE{\psi}{\varphi} & = \PSUNTIL {(\varphi \wedge \neg \psi)}{\neg \psi} \\ 
\WHEN{\psi}{\varphi} & =\BEFORE{(\neg\varphi \wedge \psi)}{(\varphi \wedge \psi)}  & \qquad \PWHEN{\psi}{\varphi} & =\PBEFORE{(\neg\varphi \wedge \psi)}{(\varphi \wedge \psi)} \\
\SWHEN{\psi}{\varphi} & =\SUNTIL{(\varphi \wedge \psi)}{\neg \psi}  & \qquad \PSWHEN{\psi}{\varphi} & =\PSUNTIL{(\varphi \wedge \psi)}{\neg \psi} \\
\end{alignat*}

\noindent For example, $\UNTIL{\psi}{\varphi}$ is the \emph{weak until} operator that can be alternatively defined as $\UNTIL{\psi}{\varphi}:=\SUNTIL{\psi}{\varphi} \vee \ALWAYS{\varphi}$, \ie the event $\psi$ that is awaited for need not hold in the future. To distinguish weak and strong operators, the strong variants of a temporal operator are underlined in this paper (as done above).

\subsection{\texorpdfstring{$\omega$-Automata}{omega-Automata}}

\begin{definition}[$\omega$-Automata ]
A \emph{$\omega$-automaton} $\mA=(\cQ,\Sigma,\cI,\cR,\cA)$ over the alphabet $\Sigma$ is given by a finite set of states $\cQ$, a set $\cI$ of initial states, a transition relation $\cR \subseteq \cQ \times \Sigma \times \cQ$ and an acceptance condition $\cA:\cQ^\omega\rightarrow \{\false,\true\}$.
\end{definition}

Given an automaton $\mA=(\cQ,\Sigma,\cI,\cR,\cA)$ and an infinite word $\alpha=a_0,a_1,\dots$ over $\Sigma$. Each infinite word $\beta=q_0,q_1,\dots$ with $q_0\in \cI$ and $q_{i+1} \in \delta(q_i,\alpha_i)$ for $i>0$ is called a run of $\alpha$ through $\mA$. The run is \emph{accepting} if $\cA(\beta)=\true$. We say that $\mA$ accepts $\alpha$ whenever an accepting run of $\alpha$ through $\mA$ exists.

Using standard terminology, we say that $\mA$ is \emph{deterministic}, if exactly one initial state exists and for each $q \in \cQ$ and each input $\sigma \in \Sigma$ there exists exactly one $s' \in \cS$ with  $(s,\sigma,s') \in \cR$. In that case we write $\mA=(\cQ,\Sigma,q_0,\delta,\cA)$ with an initial state $q_0$ and a deterministic transition function $\delta:\cQ \times \Sigma \rightarrow \cQ$.
 
In the following, we assume that $\cQ=2^\Var{}$ for a set $\Var{}$ of state variables. Moreover, we assume sets $X$ and $Y$ of input and output variables that form the inputs $\cX=2^X$ and outputs $\cY=2^Y$ of the system such that $\Sigma=\cX \times \cY$. Having this view, we define a state set $\cQ_\varphi$ to contain exactly those states where the propositional encoding of the state variables $\Var{}$ satisfy $\varphi$. Thus, we can conveniently define acceptance conditions by \LTL{} specifications.

\subsection{Classical Acceptance Conditions}

In the past, several kinds of acceptance conditions have been proposed and their different expressivenesses have been studied in depth. In particular, the following acceptance conditions have been considered \cite{Wagn79,Thom90a,Schn03}.

\begin{itemize}
\item A run is accepted by a safety condition $\ALWAYS{\varphi}$ if the run exclusively runs through the set $\cQ_\varphi$.
\item A run is accepted by a liveness condition $\EVENTUAL{\varphi}$ if the run visits at least one state of the set $\cQ_\varphi$ at least once.
\item A run is accepted by a prefix\footnote{These condititions are also called Staiger-Wagner or obligation conditions.} condition
$\bigwedge_i \left(\ALWAYS{\varphi_i} \vee \EVENTUAL{\psi_i}\right)$  if for all $i$ either the run exclusively runs through the set $\cQ_{\varphi_i}$ or visits $\cQ_{\psi_i}$ at least once.
\item A run is accepted by a Büchi condition $\ALWAYS{\EVENTUAL{\varphi}}$  if the run visits at least one state of the set $\cQ_{\varphi}$ infinitely often.
\item A run is accepted by a co-Büchi condition $\EVENTUAL{\ALWAYS{\varphi}}$  if the run visits only states of the set $\cQ_\varphi$ infinitely often.
\item Finally, a run is accepted by a Streett (or reactivity) condition $\bigwedge_{i=0}^{f}\ALWAYS\EVENTUAL{\varphi_j}\vee \EVENTUAL\ALWAYS{\psi_i}$ if for all $i$ either the run visits at least one state from $\cQ_{\varphi_i}$ or the run visits only states of the set $\cQ_{\psi_i}$ infinitely often. 
\end{itemize}



\subsection{GR(1)-Specifications for \texorpdfstring{\LTL{}}{LTL} Synthesis }
The task of \LTL{} synthesis is to develop a system that controls the output variables $Y$ so that no matter how the environment chooses the input variables $X$, a \LTL{} specification is satisfied. Thus, instead of using one of the classical acceptance conditions, it is more convenient for synthesis to consider specifications of the form $\varphi \rightarrow \psi$ where $\varphi$ represents assumptions on the environment and $\psi$ represents conclusions/guarantees the system has to satisfy. In particular, Generalized Reactivity (1) acceptance \cite{BGJP07,BGJP07a,JGWB07,PiPS06} attracted some interest in the community: here the assumptions and guarantees are all Büchi conditions, \ie we seek a system satisfying the following
acceptance condition:

\begin{equation}
GR(1):=\left(\bigwedge_{i=1}^n \ALWAYS{\EVENTUAL{p_i}} \right) \rightarrow \left(\bigwedge_{j=1}^m \ALWAYS{\EVENTUAL{q_j}} \right)
\label{eq:0}
\end{equation}

\noindent The class of specifications to which the algorithms of \cite{BGJP07,BGJP07a,JGWB07,PiPS06} can be applied is much more general than the limited form presented in equation \ref{eq:0}: The algorithm can be applied to any specification of the form $\left(\bigwedge_{i=1}^n \varphi_i \right) \rightarrow \left(\bigwedge_{i=1}^m \psi_j \right)$ where each $\varphi_i$, $\psi_j$ is specified by a deterministic Büchi automaton.

\begin{definition}[\cite{KoHB09}]\label{def:GR1-Automaton}
Assume we are given $n$ deterministic Büchi automata $\mA_1^a, \dots \mA_n^a$ for the environment's assumptions and $m$ deterministic Büchi automata $\mA_1^g, \dots \mA_m^g$ for the system's guarantees with $\mA_i^a=(\cQ_i^a,\Sigma,q_{0,i}^a,\delta_i^a,\ALWAYS\EVENTUAL{p_i})$ and $\mA_j^a=(\cQ_j^a,\Sigma,q_{0,j}^a,\delta_j^a,\ALWAYS\EVENTUAL{q_j})$. Then, we define an automaton $\mA^{GR(1)}=(\cQ,\Sigma,\delta,q_0,\cA)$ as the product of all automata $\mA_i^a$ and $\mA_j^g$ where the state space is $\cQ=\cQ_1^a \times \dots \times \cQ_n^a\times \cQ_1^g \times \dots \times\cQ_m^g$, the transition function is $\delta((q_1^a,\dots q_m^g),\sigma)=(\delta_1^a(q_1^a,\sigma), \dots ,\delta_m^g(q_m^g,\sigma))$ and the initial state is $q_0=(q_{0,1}^a, \dots q_{0,m}^g)$. The acceptance condition $\cA=\left(\bigwedge_{i=1}^n \ALWAYS{\EVENTUAL{p_i}} \right) \rightarrow \left(\bigwedge_{j=1}^m \ALWAYS{\EVENTUAL{q_j}} \right)$ is a GR(1) condition.
\end{definition}
Thus, a run of $\mA^{GR(1)}$ is accepting if either all sets $\cQ_{q_j}$ are visited infinitely often or at least some set $\cQ_{p_i}$ is visited only finitely often.

\subsection{Games}
A \emph{game} $\mG=(\cQ,\Sigma,q_0,\delta,\cA)$ is a deterministic $\omega$-automaton with an input alphabet $\Sigma=\cX \times \cY$. A \emph{play} of $\mG$ is an infinite sequence of states $\pi=q_0q_1q_2 \dots \in \cQ^\omega$ where $q_{i+1}=\delta(q_i,\sigma_i)$ for $i \geq 0$. The letters $\sigma_i=(x_i,y_i)$ are successively chosen by the players: in each step, the environment first chooses $x_i$, and then the system chooses $y_i$. A play $\pi$ is won by the system if $\cA(\pi)=\true$. Otherwise, the game is won by the environment. Note that the environment cannot react to the outputs generated by the system and thus acts like a Moore machine. In contrast, the system we would like to synthesize acts like a Mealy machine. 

We solve the game, attempting to decide whether the game is winning for the environment or the system. If the environment is winning, the specification is unrealizable. If the system is winning, we synthesize a winning strategy (which is essentially a Mealy automaton) using the algorithms given in \cite{BGJP07,BGJP07a,JGWB07,PiPS06}. 

\noindent Previous works regarding the synthesis with respect to GR(1)-synthesis had to manually generate the deterministic automata. In this paper, we show how to automatically obtain deterministic Büchi automata from a fragment of \LTL{} using the well-known Breakpoint construction. This fragment of \LTL{} is a natural fragment of \LTL{} embedded in the well-known temporal-logic hierarchy \cite{MaPn87c,ChMP92,MaPn90,MaPn91,Schn01b,Schn03}.

\section{Temporal Logic vs. Automaton Hierarchy}

\subsection{The Automaton Hierarchy}

\noindent The classical acceptance conditions, i.e., safety, guarantee/liveness, fairness/response/Büchi, persistence/co-Büchi properties, define the corresponding automaton classes ${\sf (N)Det}_{\ALWAYS{}}$, ${\sf (N)Det}_{\EVENTUAL{}}$, ${\sf (N)Det}_{\ALWAYS{\EVENTUAL{}}}$, and ${\sf (N)Det}_{\EVENTUAL{\ALWAYS{}}}$, respectively. Moreover, their boolean closures can be represented by the automaton classes ${\sf (N)Det}_{\sf Prefix}$ and ${\sf (N)Det}_{\sf Streett}$ whose acceptance conditions have the forms $\bigwedge_{j=0}^{f} \ALWAYS{\varphi_j} \vee \EVENTUAL{\psi_j}$ and $\bigwedge_{j=0}^{f}\ALWAYS\EVENTUAL{\varphi_j}\vee \EVENTUAL\ALWAYS{\psi_j}$, respectively.

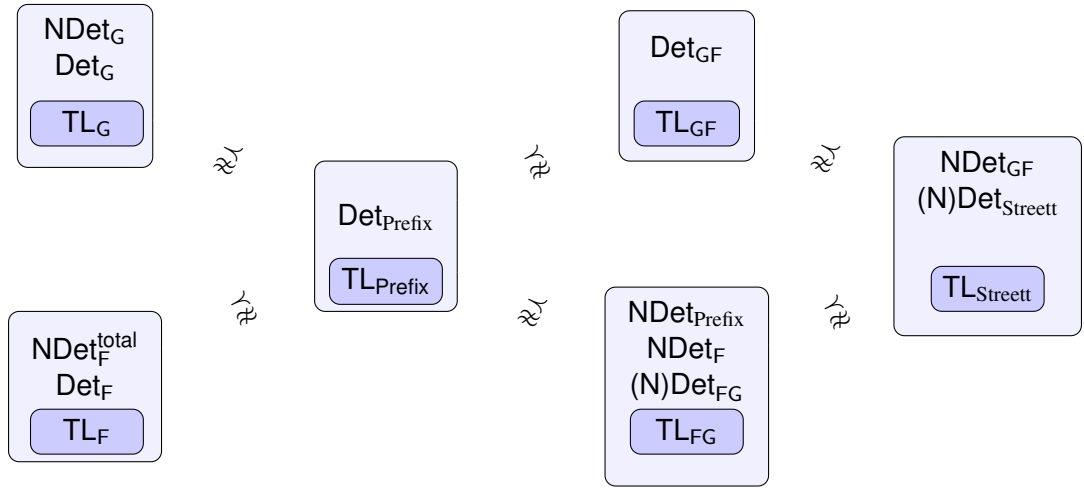
\begin{figure}
\begin{centering}
\tikzstyle{tikblock} = [draw, fill=blue!6, rounded corners]
\begin{tikzpicture}[shorten >=1pt,node distance=2cm,auto]
\node[tikblock,minimum height=2cm,minimum width=1.8cm] (G) at (0,5) {$\begin{array}{c}
						\text{\sf NDet}_{\sf \ALWAYS{}}\\
						\text{\sf Det}_{\sf \ALWAYS{}}
						\\\vspace{\baselineskip} 
						\\\vspace{\baselineskip} 
					\end{array}$};
\node[tikblock,minimum height=2cm,minimum width=1.8cm] (F) at (0,1)  {$\begin{array}{c}
 							\text{\sf NDet}_{{\sf \EVENTUAL{}}}^\text{\sf total}\\
							\text{\sf Det}_{{\sf \EVENTUAL{}}}
								\\\vspace{1\baselineskip} 
							\end{array}$};
\node[tikblock,minimum height=2cm,minimum width=1.8 cm] (Prefix) at (4,3) {$\begin{array}{c}
							\text{\sf Det}_\text{Prefix}
								\\\vspace{1.5\baselineskip}
							\end{array}$};
\node[tikblock,minimum height=2cm,minimum width=1.8cm] (GF) at (8,5) {$\begin{array}{c}
 							\text{\sf Det}_{{\ALWAYS\EVENTUAL{}}}
								\\\vspace{\baselineskip} 
								\\\vspace{\baselineskip} 
							\end{array}$};
\node[tikblock,minimum height=2cm] (FG) at (8,1) {$\begin{array}{c}
 							\text{\sf NDet}_\text{Prefix}\\
 							\text{\sf NDet}_{\EVENTUAL{}}\\
 							\text{\sf (N)Det}_{\EVENTUAL{\ALWAYS{}}}
						    \\\vspace{\baselineskip} 
							\\\vspace{\baselineskip}
							\end{array}$};
\node[tikblock,minimum height=2cm] (Streett) at (12,3) {$\begin{array}{c}
 							\text{\sf NDet}_{{\ALWAYS\EVENTUAL{}}} \\
							\text{\sf (N)Det}_{\text{Streett}} \\
								\\\vspace{1\baselineskip}
								\\\vspace{1\baselineskip}\\
							\end{array}$};
							
\node[tikblock,fill=blue!20,minimum width=1.5cm] at (0.025,4.5) {$\text{\sf TL}_{\ALWAYS{}}$};
\node[tikblock,fill=blue!20,minimum width=1.5cm] at (0.025,0.4) {$\text{\sf TL}_{\EVENTUAL{}}$};
\node[tikblock,fill=blue!20,minimum width=1.5cm] at (4,2.4) {$\text{\sf TL}_\text{\sf Prefix}$};
\node[tikblock,fill=blue!20,minimum width=1.5cm] at (12,2.3) {$\text{\sf TL}_{\text{Streett}}$};
\node[tikblock,fill=blue!20,minimum width=1.4cm] at (8,4.5) {$\text{\sf TL}_{\ALWAYS{\EVENTUAL{}}}$};
\node[tikblock,fill=blue!20,minimum width=1.5cm] at (8,0.4) {$\text{\sf TL}_{\EVENTUAL{\ALWAYS{}}}$};

\path[->,color=white] (G) edge node [sloped,left=-0.3cm] {\color{black} $\lsexpress{}{}$} (Prefix)
	     (F) edge node [sloped,right=-0.3cm] {\color{black} $\lsexpress{}{}$} (Prefix)
	     (Prefix) edge node [sloped,right=-0.4cm] {\color{black} $\lsexpress{}{}$} (GF)
	     (Prefix) edge node [sloped,left=-0.4cm] {\color{black}  $\lsexpress{}{}$} (FG)	
	     (GF) edge node [sloped,left=-0.4cm] {\color{black} $\lsexpress{}{}$} (Streett)
	     (FG) edge node [sloped,right=-0.3cm] {\color{black} $\lsexpress{}{}$} (Streett)	;		
\end{tikzpicture}
\caption{(Borel) Hierarchy of $\omega$-Automata and Temporal Logic}
\label{fig:automata_hierarchy}
\end{centering}
\end{figure}

The expressiveness of these classes is illustrated in Figure~\ref{fig:automata_hierarchy}, where $\lseqexpress{\cC_1}{\cC_2}$ means that for any automaton in $\cC_1$, there is an equivalent one in $\cC_2$. Moreover, we define $\eqexpress{\cC_1}{\cC_2} := \lseqexpress{\cC_1}{\cC_2} \wedge \lseqexpress{\cC_2}{\cC_1}$ and $\lsexpress{\cC_1}{\cC_2} := \lseqexpress{\cC_1}{\cC_2} \wedge \neg (\eqexpress{\cC_1}{\cC_2})$. As can be seen, the hierarchy consists of six different classes, and each class has a deterministic representative. 

\subsection{The Temporal Logic Hierarchy}

In \cite{ChMP92,Schn01b,Schn03}, corresponding hierarchies for temporal logics have been defined. Following \cite{Schn01b,Schn03}, we define the hierarchy of temporal logic formulas syntactically by the grammar rules of Fig.~\ref{fig:grammar}:

\begin{figure*}[!th] \[
\begin{array}{|c|c|}
\hline
	\begin{array}{ll}
	P_{\ALWAYS{}} ::= &
		\Var{\Sigma}
		\mid \neg P_{\EVENTUAL{}}  
		\mid P_{\ALWAYS{}}\wedge P_{\ALWAYS{}} 
		\mid P_{\ALWAYS{}}\vee P_{\ALWAYS{}} \\ &
		\mid \PNEXT{P_{\ALWAYS{}}} 
		\mid \PUNTIL{P_{\ALWAYS{}}}{P_{\ALWAYS{}}}  \\ &
		\mid \PSNEXT{P_{\ALWAYS{}}} 
		\mid \PSUNTIL{P_{\ALWAYS{}}}{P_{\ALWAYS{}}}\\ &
		\mid \NEXT{P_{\ALWAYS{}}} 
		\mid \UNTIL{P_{\ALWAYS{}}}{P_{\ALWAYS{}}} \\
	\end{array} &
	\begin{array}{ll}
	P_{\EVENTUAL{}} ::= &
		\Var{\Sigma}
		\mid \neg P_{\ALWAYS{}}  
		\mid P_{\EVENTUAL{}}\wedge P_{\EVENTUAL{}} 
		\mid P_{\EVENTUAL{}}\vee P_{\EVENTUAL{}} \\ &
		\mid \PNEXT{P_{\EVENTUAL{}}} 
		\mid \PUNTIL{P_{\EVENTUAL{}}}{P_{\EVENTUAL{}}}\\ &
		\mid \PSNEXT{P_{\EVENTUAL{}}} 
		\mid \PSUNTIL{P_{\EVENTUAL{}}}{P_{\EVENTUAL{}}} \\ &
		\mid \NEXT{P_{\EVENTUAL{}}} 
		\mid \SUNTIL{P_{\EVENTUAL{}}}{P_{\EVENTUAL{}}} \\
	\end{array}
\\
\hline
\multicolumn{2}{|c|}{%
\begin{array}{ll}
P_{\sf Prefix} ::= &
	     P_{\ALWAYS{}}  
	\mid P_{\EVENTUAL{}} 
	\mid \neg P_{\sf Prefix}
	\mid P_{\sf Prefix} \wedge P_{\sf Prefix}
	\mid P_{\sf Prefix} \vee P_{\sf Prefix}
\end{array}}
\\\hline
	\begin{array}{ll}
	P_{\ALWAYS{\EVENTUAL{}}} ::= & P_{\sf Prefix}  \\ &
		\mid \neg P_{\EVENTUAL{\ALWAYS{}}}
		\mid P_{\ALWAYS{\EVENTUAL{}}}\wedge P_{\ALWAYS{\EVENTUAL{}}} 
		\mid P_{\ALWAYS{\EVENTUAL{}}}\vee P_{\ALWAYS{\EVENTUAL{}}} \\ &
		\mid \PNEXT{P_{\ALWAYS{\EVENTUAL{}}}} 
		\mid \PSNEXT{P_{\ALWAYS{\EVENTUAL{}}}} 
		\mid \NEXT{P_{\ALWAYS{\EVENTUAL{}}}} \\ &
		\mid \PUNTIL{P_{\ALWAYS{\EVENTUAL{}}}}{P_{\ALWAYS{\EVENTUAL{}}}}  
		\mid \PSUNTIL{P_{\ALWAYS{\EVENTUAL{}}}}{P_{\ALWAYS{\EVENTUAL{}}}} \\ &
		\mid  \UNTIL{P_{\ALWAYS{\EVENTUAL{}}}}{P_{\ALWAYS{\EVENTUAL{}}}} 
		\mid \SUNTIL{P_{\EVENTUAL{}}}{P_{\ALWAYS{\EVENTUAL{}}}}
	\end{array} &
	\begin{array}{ll}
	P_{\EVENTUAL{\ALWAYS{}}} ::= & P_{\sf Prefix}  \\ &
		\mid \neg P_{\ALWAYS{\EVENTUAL{}}}
		\mid P_{\EVENTUAL{\ALWAYS{}}}\wedge P_{\EVENTUAL{\ALWAYS{}}} 
		\mid P_{\EVENTUAL{\ALWAYS{}}}\vee P_{\EVENTUAL{\ALWAYS{}}} \\ &
		\mid \PNEXT{P_{\EVENTUAL{\ALWAYS{}}}} 
		\mid \NEXT{P_{\EVENTUAL{\ALWAYS{}}}} 
		\mid \PSNEXT{P_{\EVENTUAL{\ALWAYS{}}}} \\ &
		\mid    \PUNTIL{P_{\EVENTUAL{\ALWAYS{}}}}{P_{\EVENTUAL{\ALWAYS{}}}} 
		\mid   \PSUNTIL{P_{\EVENTUAL{\ALWAYS{}}}}{P_{\EVENTUAL{\ALWAYS{}}}} \\ &
		\mid    \SUNTIL{P_{\EVENTUAL{\ALWAYS{}}}}{P_{\EVENTUAL{\ALWAYS{}}}} 
		\mid     \UNTIL{P_{\EVENTUAL{\ALWAYS{}}}}{P_{\ALWAYS{}}}
	\end{array}
\\
\hline
\multicolumn{2}{|c|}{%
	\begin{array}{ll}
	P_{\sf Streett} ::= &
	             P_{\ALWAYS{\EVENTUAL{}}}  
		\mid P_{\EVENTUAL{\ALWAYS{}}} 
		\mid \neg P_{\sf Streett}
		\mid P_{\sf Streett} \wedge P_{\sf Streett}
		\mid P_{\sf Streett} \vee P_{\sf Streett}
\end{array}
}
\\
\hline
\end{array}
\]
\caption{Syntactic Characterizations of the Classes of the Temporal Logic Hierarchy}
\label{fig:grammar}
\end{figure*}

\begin{definition}[Temporal Logic Classes] \label{temp_borel_1_def}
For $\kappa\in\{\ALWAYS{}$, $\EVENTUAL{}$, ${\sf Prefix}$, $\EVENTUAL{\ALWAYS{}}$, $\ALWAYS{\EVENTUAL{}}$, ${\sf Streett} \}$, we define the logics ${\sf TL}_\kappa$ by the grammars given in Fig.~\ref{fig:grammar}, where ${\sf TL}_\kappa$ is the set of formulas that can be derived from the nonterminal $P_\kappa$ ($\Var{\Sigma}$ represents any variable $v\in\Var{\Sigma}$).
\end{definition}

\noindent Typical safety conditions like $\ALWAYS{\varphi}$ or $\ALWAYS{\UNTIL{b}{a}}$ that state that something bad never happens, are contained in ${\sf TL}_\ALWAYS{}$. Liveness conditions like $\EVENTUAL{\varphi}$ are contained in ${\sf TL}_\EVENTUAL{}$. Finally, fairness conditions like $\ALWAYS\EVENTUAL{\varphi}$ that demand that something good infinitely often happens, are contained in ${\sf TL}_{\ALWAYS\EVENTUAL{}}$ while stabilization/persistence properties like $\EVENTUAL\ALWAYS{\varphi}$ that demand that after a finite interval, nothing bad happens are contained in ${\sf TL}_{\EVENTUAL\ALWAYS{}}$.

\subsection{Relating the Temporal Logic and the Automata Hierarchy}

In \cite{Schn01b,Schn03} several translation procedures are given to translate formulas from ${\sf TL}_{\kappa}$ to equivalent ${\sf (N)Det}_\kappa$ automata. In particular, the following is an important result:

\begin{theorem}[Temporal Logic and Automaton Hierarchy] \label{borel_thm}
Given a formula $\Phi \in {\sf TL}_\kappa$, we can construct a deterministic $\omega$-automaton $\mA=(\Pot{Q},\cI,\cR,\lambda,\cA)$ of the class ${\sf Det}_\kappa$ in time $O(2^{\card{\Phi}})$ with $\card{Q}\leq 2^{\card{\Phi}}$ state variables. Therefore, $\mA=(\Pot{Q},\cI,\cR,\lambda,\cA)$ is a symbolic representation of a deterministic automaton with $O(2^{2^{\card{\Phi}}})$ states.
\end{theorem}

\noindent The above results are already proved in detail in \cite{Schn03}, where translation procedures from ${\sf TL}_\kappa$ to ${\sf NDet}_\kappa$ have been constructed. Moreover, it has been shown in \cite{Schn03} that the subset construction can be used to determinize the automata that stem from the classes ${\sf TL}_{\ALWAYS{}}$ and ${\sf TL}_{\EVENTUAL{}}$ and that the Miyano-Hayashi breakpoint construction is sufficient to determinize the automata that stem from the translation of formulas from ${\sf TL}_{\EVENTUAL{\ALWAYS{}}}$ and ${\sf TL}_{\ALWAYS{\EVENTUAL{}}}$. Since ${\sf TL}_{\sf Prefix}$ and ${\sf TL}_{\sf Streett}$ are the boolean closures of ${\sf TL}_{\ALWAYS{}}\cup{\sf TL}_{\EVENTUAL{}}$ and ${\sf TL}_{\EVENTUAL{\ALWAYS{}}}\cup{\sf TL}_{\ALWAYS{\EVENTUAL{}}}$, respectively, the remaining results for ${\sf TL}_{\sf Prefix}$ and ${\sf TL}_{\sf Streett}$ follow from the boolean combinations of ${\sf Det}_{\ALWAYS{}}/{\sf Det}_{\EVENTUAL{}}$ and ${\sf Det}_{\EVENTUAL{\ALWAYS{}}}/{\sf Det}_{\ALWAYS{\EVENTUAL{}}}$, respectively. 

The final step consists of computing the boolean closure of the acceptance conditions. To this end, it is shown in \cite{Schn03} how arbitrary boolean combinations of $\ALWAYS{\varphi}$ and $\EVENTUAL{\varphi}$ with propositional formulas $\varphi$ are translated to equivalent ${\sf Det}_{\sf Prefix}$ automata, and analogously, how arbitrary boolean combinations of $\ALWAYS{\EVENTUAL{\varphi}}$ and $\EVENTUAL{\ALWAYS{\varphi}}$ with propositional formulas $\varphi$ are translated to equivalent ${\sf Det}_{\sf Streett}$ automata.

\begin{figure*}[!th] \[
\begin{array}{|c|c|}
\hline
	\begin{array}{ll}
	P_{\ALWAYS{}} ::= &
		\Var{\Sigma}
		\mid \neg P_{\EVENTUAL{}}  
		\mid P_{\ALWAYS{}}\wedge P_{\ALWAYS{}} 
		\mid P_{\ALWAYS{}}\vee P_{\ALWAYS{}} \\ &
		\mid \PNEXT{P_{\ALWAYS{}}} 
		\mid \PUNTIL{P_{\ALWAYS{}}}{P_{\ALWAYS{}}}  \\ &
		\mid \PSNEXT{P_{\ALWAYS{}}} 
		\mid \PSUNTIL{P_{\ALWAYS{}}}{P_{\ALWAYS{}}}\\ &
		\mid \NEXT{P_{\ALWAYS{}}} 
		\mid \UNTIL{P_{\ALWAYS{}}}{P_{\ALWAYS{}}} \\
	\end{array} &
	\begin{array}{ll}
	P_{\EVENTUAL{}} ::= &
		\Var{\Sigma}
		\mid \neg P_{\ALWAYS{}}  
		\mid P_{\EVENTUAL{}}\wedge P_{\EVENTUAL{}} 
		\mid P_{\EVENTUAL{}}\vee P_{\EVENTUAL{}} \\ &
		\mid \PNEXT{P_{\EVENTUAL{}}} 
		\mid \PUNTIL{P_{\EVENTUAL{}}}{P_{\EVENTUAL{}}}\\ &
		\mid \PSNEXT{P_{\EVENTUAL{}}} 
		\mid \PSUNTIL{P_{\EVENTUAL{}}}{P_{\EVENTUAL{}}} \\ &
		\mid \NEXT{P_{\EVENTUAL{}}} 
		\mid \SUNTIL{P_{\EVENTUAL{}}}{P_{\EVENTUAL{}}} \\
	\end{array}
\\
\hline
\multicolumn{2}{|c|}{%
\begin{array}{ll}
P_{\sf Prefix} ::= &
	     P_{\ALWAYS{}}  
	\mid P_{\EVENTUAL{}} 
	\mid \neg P_{\sf Prefix}
	\mid P_{\sf Prefix} \wedge P_{\sf Prefix}
	\mid P_{\sf Prefix} \vee P_{\sf Prefix}
\end{array}}
\\\hline
	\begin{array}{ll}
	P_{\ALWAYS{\EVENTUAL{}}} ::= & P_{\sf Prefix}  \\ &
		\mid \neg P_{\EVENTUAL{\ALWAYS{}}}
		\mid P_{\ALWAYS{\EVENTUAL{}}}\wedge P_{\ALWAYS{\EVENTUAL{}}} 
		\mid P_{\ALWAYS{\EVENTUAL{}}}\vee P_{\ALWAYS{\EVENTUAL{}}} \\ &
		\mid \PNEXT{P_{\ALWAYS{\EVENTUAL{}}}} 
		\mid \PSNEXT{P_{\ALWAYS{\EVENTUAL{}}}} 
		\mid \NEXT{P_{\ALWAYS{\EVENTUAL{}}}} \\ &
		\mid \PUNTIL{P_{\ALWAYS{\EVENTUAL{}}}}{P_{\ALWAYS{\EVENTUAL{}}}}  
		\mid \PSUNTIL{P_{\ALWAYS{\EVENTUAL{}}}}{P_{\ALWAYS{\EVENTUAL{}}}} \\ &
		\mid  \UNTIL{P_{\ALWAYS{\EVENTUAL{}}}}{P_{\ALWAYS{\EVENTUAL{}}}} 
		\mid \SUNTIL{P_{\EVENTUAL{}}}{P_{\ALWAYS{\EVENTUAL{}}}}
	\end{array} &
	\begin{array}{ll}
	P_{\EVENTUAL{\ALWAYS{}}} ::= & P_{\sf Prefix}  \\ &
		\mid \neg P_{\ALWAYS{\EVENTUAL{}}}
		\mid P_{\EVENTUAL{\ALWAYS{}}}\wedge P_{\EVENTUAL{\ALWAYS{}}} 
		\mid P_{\EVENTUAL{\ALWAYS{}}}\vee P_{\EVENTUAL{\ALWAYS{}}} \\ &
		\mid \PNEXT{P_{\EVENTUAL{\ALWAYS{}}}} 
		\mid \NEXT{P_{\EVENTUAL{\ALWAYS{}}}} 
		\mid \PSNEXT{P_{\EVENTUAL{\ALWAYS{}}}} \\ &
		\mid    \PUNTIL{P_{\EVENTUAL{\ALWAYS{}}}}{P_{\EVENTUAL{\ALWAYS{}}}} 
		\mid   \PSUNTIL{P_{\EVENTUAL{\ALWAYS{}}}}{P_{\EVENTUAL{\ALWAYS{}}}} \\ &
		\mid    \SUNTIL{P_{\EVENTUAL{\ALWAYS{}}}}{P_{\EVENTUAL{\ALWAYS{}}}} 
		\mid     \UNTIL{P_{\EVENTUAL{\ALWAYS{}}}}{P_{\ALWAYS{}}}
	\end{array}
\\
\hline
	\begin{array}{ll}
	P_{\sf Assume} ::= &
		P_{\ALWAYS{\EVENTUAL{}}}
		\mid P_{\sf Assume} \wedge P_{\sf Assume}
	\end{array} &
	\begin{array}{ll}
	P_{\sf Guarantee} ::= &
		P_{\ALWAYS{\EVENTUAL{}}}
		\mid P_{\sf Guarantee} \wedge P_{\sf Guarantee}
	\end{array} \\
\hline
	 \multicolumn{2}{|c|}{%
	\begin{array}{ll}
	P_{\sf GR(1)} ::= P_{\sf Assume} \rightarrow P_{\sf Assert}
	\end{array}}\\
\hline
\end{array}
\]
\caption{A \LTL{} Fragment for GR(1)-Synthesis}
\label{fig:grammar2}
\end{figure*}

\section{A \texorpdfstring{\LTL{}}{LTL} Fragment for GR(1)-Synthesis}

Using the previously mentioned temporal logic hierarchy, we define a fragment of \LTL{} that can be easily translated to a set of deterministic Büchi automata for the assumptions and a set of deterministic Büchi automata for the guarantees (Figure \ref {fig:grammar2}). 

As can be seen, our \LTL{} fragment is naturally embedded in the temporal logic hierarchy. The formulas that syntactically belong to our \LTL{} fragment are those formulas that are derived from the nonterminal $P_{\sf GR(1)}$, thus, these are implications of formulas that are derived from the nonterminals $P_{\sf Assume}$ and $P_{\sf Assert}$, respectively, which are both conjunctions of ${\sf TL}_{\ALWAYS\EVENTUAL{}}$-formulas. 

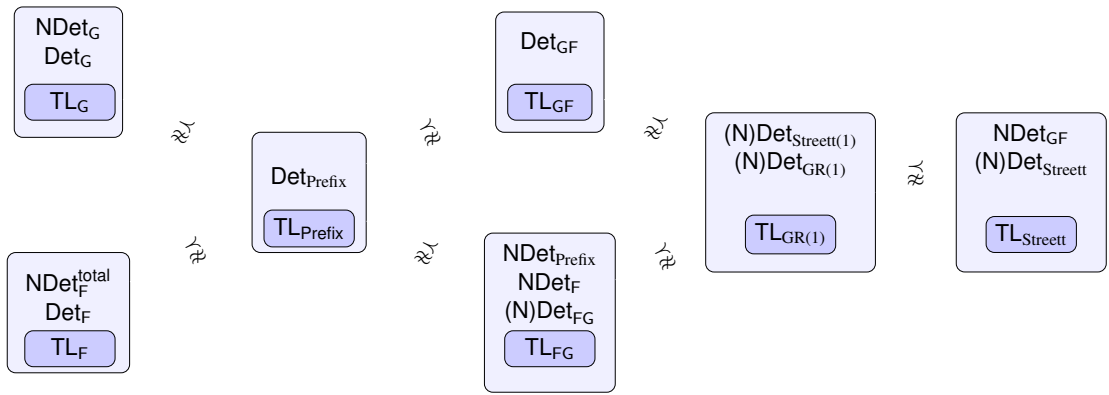
\begin{figure}
\begin{centering}
\scalebox{0.8}{
\tikzstyle{tikblock} = [draw, fill=blue!6, rounded corners]
\begin{tikzpicture}[shorten >=1pt,node distance=2cm,auto]
    	 \node[tikblock,minimum height=2cm,minimum width=1.8cm] (G) at (0,5) {$\begin{array}{c}
	 							\text{\sf NDet}_{\sf \ALWAYS{}}\\
	 							\text{\sf Det}_{\sf \ALWAYS{}}
									\\\vspace{\baselineskip} 
									\\\vspace{\baselineskip} 
								\end{array}$};
	 \node[tikblock,minimum height=2cm,minimum width=1.8cm] (F) at (0,1)  {$\begin{array}{c}
	 							\text{\sf NDet}_{{\sf \EVENTUAL{}}}^\text{\sf total}\\
								\text{\sf Det}_{{\sf \EVENTUAL{}}}
									\\\vspace{1\baselineskip} 
								\end{array}$};
	\node[tikblock,minimum height=2cm,minimum width=1.8 cm] (Prefix) at (4,3) {$\begin{array}{c}
								\text{\sf Det}_\text{Prefix}
									\\\vspace{1.5\baselineskip}
								\end{array}$};
	\node[tikblock,minimum height=2cm,minimum width=1.8cm] (GF) at (8,5) {$\begin{array}{c}
	 							\text{\sf Det}_{{\ALWAYS\EVENTUAL{}}}
									\\\vspace{\baselineskip} 
									\\\vspace{\baselineskip} 
								\end{array}$};
	\node[tikblock,minimum height=2cm] (FG) at (8,1) {$\begin{array}{c}
	 							\text{\sf NDet}_\text{Prefix}\\
	 							\text{\sf NDet}_{\EVENTUAL{}}\\
	 							\text{\sf (N)Det}_{\EVENTUAL{\ALWAYS{}}}
							    \\\vspace{\baselineskip} 
								\\\vspace{\baselineskip}
								\end{array}$};
	\node[tikblock,minimum height=2cm] (Streett1) at (12,3) {$\begin{array}{c}
	 							\text{\sf (N)Det}_{\text{Streett(1)}} \\
								\text{\sf (N)Det}_{\text{GR(1)}} \\
								\\\vspace{1\baselineskip}
								\\\vspace{1\baselineskip}\\
							
								\end{array}$};
	\node[tikblock,minimum height=2cm] (Streett) at (16,3) {$\begin{array}{c}
	 							\text{\sf NDet}_{{\ALWAYS\EVENTUAL{}}} \\
								\text{\sf (N)Det}_{\text{Streett}} \\
									\\\vspace{1\baselineskip}
									\\\vspace{1\baselineskip}\\
								\end{array}$};
								
	\node[tikblock,fill=blue!20,minimum width=1.5cm] at (0.025,4.5) {$\text{\sf TL}_{\ALWAYS{}}$};
	\node[tikblock,fill=blue!20,minimum width=1.5cm] at (0.025,0.4) {$\text{\sf TL}_{\EVENTUAL{}}$};
	\node[tikblock,fill=blue!20,minimum width=1.5cm] at (4,2.4) {$\text{\sf TL}_\text{\sf Prefix}$};
	\node[tikblock,fill=blue!20,minimum width=1.5cm] at (16,2.3) {$\text{\sf TL}_{\text{Streett}}$};
	
	\node[tikblock,fill=blue!20,minimum width=1.5cm] at (12,2.3) {$\text{\sf TL}_{\text{GR(1)}}$};
	\node[tikblock,fill=blue!20,minimum width=1.4cm] at (8,4.5) {$\text{\sf TL}_{\ALWAYS{\EVENTUAL{}}}$};
	\node[tikblock,fill=blue!20,minimum width=1.5cm] at (8,0.4) {$\text{\sf TL}_{\EVENTUAL{\ALWAYS{}}}$};
	
	\path[->,color=white] (G) edge node [sloped,left=-0.3cm] {\color{black} $\lsexpress{}{}$} (Prefix)
		     (F) edge node [sloped,right=-0.3cm] {\color{black} $\lsexpress{}{}$} (Prefix)
		     (Prefix) edge node [sloped,right=-0.4cm] {\color{black} $\lsexpress{}{}$} (GF)
		     (Prefix) edge node [sloped,left=-0.4cm] {\color{black}  $\lsexpress{}{}$} (FG)	
		     (GF) edge node [sloped,left=-0.4cm] {\color{black} $\lsexpress{}{}$} (Streett1)
		     (FG) edge node [sloped,right=-0.3cm] {\color{black} $\lsexpress{}{}$} (Streett1)			
		     (Streett1) edge node {\color{black} $\lsexpress{}{}$} (Streett)	;		
		     
\end{tikzpicture}
}
\end{centering}
\caption{(Borel) Hierarchy of $\omega$-Automata and Temporal Logic with GR(1)}
\label{fig:automata_hierarchy2}
\end{figure}

\noindent Concerning the automata hierarchy, we can translate these formulas to automata with a GR(1)-acceptance condition, i.e. a generalization of a Streett(1) condition. In \cite{BCGH10}, it is shown that a GR(1)-condition can be equivalently expressed by a Streett(1)-condition, \ie a Streett condition with only one acceptance pair. Hence, we obtain the "enriched" automata hierarchy shown in Figure~\ref{fig:automata_hierarchy2} together with the following corollary that easily follows from Theorem~\ref{borel_thm}:

\begin{corollary}
Given a $P_{\sf GR(1)}$-formula of the form $\Phi=\left(\varphi_1\wedge\ldots\wedge\varphi_n\right) \rightarrow \left(\psi_1 \wedge  \ldots\wedge\psi_m\right)$, we can compute $n$ deterministic Büchi automata $\cA_{\varphi_1}^a, \ldots \cA_{\varphi_n}^a$ and $m$ deterministic Büchi automata $\cA_{\psi_1}^g, \ldots \cA_{\psi_n}^g$ such that $\cA_{\varphi_i}$ ($\cA_{\psi_j}$) is initially equivalent to $\varphi_i$ (resp. $\psi_j$). Hence the GR(1)-automaton obtained from those automata according to Definition \ref{def:GR1-Automaton} is initially equivalent to $\Phi$.
\end{corollary}

\section{Experiments}

In our previous work, we had already implemented a toolset Averest \cite{Schn09} whose inputs are programs written in the Esterel-like synchronous programming language Quartz \cite{Schn09}. Averest compiles the synchronous programs to guarded actions which can be used in turn to generate sequential and concurrent software, hardware or symbolic transition relations for formal verification. Specifications can be given in various temporal logics and the $\mu$-calculus. Averest provides a lot of translations from temporal logic to either $\omega$-automata or directly to the $\mu$-calculus (see \cite{Schn03} for these translations).

\begin{figure}
\includegraphics[scale=0.65]{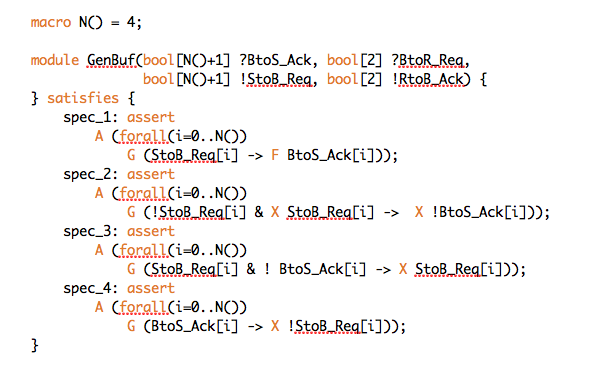}
\caption{An Example Quartz File with a GR(1) Specification having only Assertions}
\label{fig:example_spec}	
\end{figure}

For this paper, we implemented an additional tool Quartz2Marduk that takes as input a set of \LTL{} formulas that represent assumptions and assertions/guarantees of a GR(1) specification (see example shown in Figure~\ref{fig:example_spec}). We then check whether these specifications belong to the class that can be used for GR(1)-synthesis. If so, we automatically generate deterministic automata that are equivalent to the specification. The automata are automatically minimized using a form of delayed simulation \cite{Frit05b} and are afterwards used to generate a file as input to the Marduk\footnote{Actually, our current implementation generates an Anzu \cite{JGWB07} file and we use a tool included with Marduk to translate this Anzu file to a Marduk file.} tool \cite{BCGH10a}. Marduk is a re-implementation of Anzu \cite{JGWB07} with some new features.
It is basically a BDD-based implementation of the algorithm given in \cite{PiPS06}.

Included with Marduk came two case studies that are described in \cite{BGJP07,BGJP07a,JGWB07}. The first case study is the GenBuf example that is used asa tutorial in IBMs RuleBase system. The second example is \emph{ARM's Advanced Microcontroller Bus Architecture (AMBA)} which defines the \emph{Advanced High performance Bus (AHB)}, an on-chip communication standard that connects devices like processor cores, caches and DMA arbiters.

In \cite{BGJP07,BGJP07a,JGWB07} temporal logic specifications for those case studies are given along with some hints how deterministic automata for these specifications can be manually obtained. Marduk came with an input file that already contained those manually generated deterministic automata. In our tool, all we had to do is to simply write down the temporal logic specifications given in \cite{BGJP07,BGJP07a,JGWB07} and compile it to a Marduk input file.

After having compiled the Marduk input files, we ran Marduk with dynamic variable ordering enabled, leaving the other options untouched. The results of our experiments is given in table~\ref{fig:experiments}. The first column given there is the name of the case study, the second column is the time (in seconds) our tool needed to perform determinization. The third column lists the number of state variables that where generated by our tool and the manual generated deterministic automata.The next column lists the number of BDD Nodes for the generated strategy. Finally, the last column lists the runtime of Marduk for the automatically generated automata and the respective time for the manually generated automata. In the table, TO means that the synthesis procedure could not be finished within 50000 seconds\footnote{We can not satisfactorily explain why the synthesis for the AMBA model needed more time for 6  masters than for 7 masters using our determinization procedure. However, the same holds for the manually generated automata where this observation can be done for 8 respectively for 9 masters. However, a similar observation was also reported in \cite{BGJP07a}.}.

\begin{figure}[ht]
\centering
\begin{tabular}{|c|c|cc|cc|cc|}
	\hline
	Model & Det (s) & \multicolumn{2}{c|}{State Vars}& \multicolumn{2}{c|}{Strategy Nodes}& \multicolumn{2}{c|}{Solve(t)} \\
	& & Auto & Manu & Auto & Manu &Auto & Manu\\
	\hline
	GenBuf 2 & 0.1 & 12 & 3&8.755&3.344& 0.86 & 0.25\\
	GenBuf 3 & 0.1 & 12 &3&19.087&4.237& 1.96 & 0.3\\
	GenBuf 4 & 0.2 & 12 &3&25.653&5.546& 2.12 & 0.63\\
	GenBuf 5 & 0.2 &12 &3&39.356&11.916& 12.88 & 1.34\\
	GenBuf6 & 0.3 &12&3&26.139&15.605& 5.61 & 2.38\\
	GenBuf7 & 0.3 &12&3&117.625&18.894& 41.92 & 3.75\\
	GenBuf8 & 0.3 &12&3&45.238&24.302& 11.24 & 5.14\\
	GenBuf9 & 0.3 &12&3&27.507&24.493& 12.7 & 7.8\\
	GenBuf10 & 0.3 &12 &3&67.879&51.605& 44.91& 25.3\\
	Amba2 & 0.6 &9&7&38.107&50.816& 3.0 & 1.97\\
	Amba3 & 1.1 &10&8&77.033&122.027& 14.4& 10.64\\
	Amba4 & 1.8 &11&9&451.456&503.622& 66.9 & 98.32\\
	Amba5 & 7.2 &12&10&1.194.190&825.294& 1221.7 & 381.34\\
	Amba6 & 19.4 &13&11&4.929.635&989.482&  46815 & 420.96\\
	Amba7 & 42.0 &14&12&2.052.871&1.037.608&  4555.2 & 904.78 \\
	Amba8 &  83.1 &15&13&TO&3.625.518& TO & 13617.19\\
	Amba9 & 403.6 &16&14&TO&1.331.441& TO & 4215.94\\
	Amba10 & 580.16 &17&15&TO&3.034.060& TO & 7325.85\\
	\hline
\end{tabular}	
\caption{Experimental Results}
\label{fig:experiments}
\end{figure}

\section{Discussion}

The GR(1)-approach is one of the most successful approaches to \LTL{} synthesis today \cite{BGJP07,BGJP07a,JGWB07} that has already found applications apart from its primary target \cite{WoTM10}. One interesting question regarding the GR(1)-synthesis approach is its good algorithmic behavior of having a cubic runtime despite the fact that many specifications can be rewritten to a deterministic automaton having a GR(1)-acceptance condition. This question has been answered in \cite{BCGH10} where it is shown that in fact an automaton with GR(1)-acceptance condition is equivalent to a Streett automaton having only one acceptance pair. 

In this article, we gave the corresponding temporal logic view: We presented a fragment of \LTL{} that is `naturally' embedded in the temporal logic hierarchy and that can be easily translated to a corresponding deterministic GR(1)-automaton. We have implemented a tool that is able to translate any formula from this fragment to a corresponding deterministic GR(1)-automaton. This is a useful improvement in the expressivity and usage of the GR(1)-approach: instead of having the need to generate deterministic automata manually, the input to our tool is a more readable \LTL{} formula.

However, this higher expressivity comes to a cost: Not too surprisingly, running Marduk on the manually generated automata took a significant smaller amount of time than on the automatically generated automata and moreover, generated smaller BDDs for the strategies. However, the manually generated automata have undergone heavy (hand-crafted) minimization steps\footnote{Compare the difference in the runtime of the Anzu tool reported in \cite{BGJP07} with the one reported in \cite{BGJP07a}.} and hence we expect that further improvements on the determinization or the minimization step of our tool could also significantly improve our results.

\section{Acknowledgements}
We would like to thank Georg Hofferek for his kind help with the tool Marduk.

\bibliographystyle{eptcs} 
\bibliography{paper.bib}

\begin{thebibliography}{10}
\providecommand{\bibitemdeclare}[2]{}
\providecommand{\urlprefix}{Available at }
\providecommand{\url}[1]{\texttt{#1}}
\providecommand{\href}[2]{\texttt{#2}}
\providecommand{\urlalt}[2]{\href{#1}{#2}}
\providecommand{\doi}[1]{doi:\urlalt{http://dx.doi.org/#1}{#1}}
\providecommand{\bibinfo}[2]{#2}

\bibitemdeclare{article}{AlTo04}
\bibitem{AlTo04}
\bibinfo{author}{R.~Alur} \& \bibinfo{author}{S.~{La Torre}}
  (\bibinfo{year}{2004}): \emph{\bibinfo{title}{Deterministic Generators and
  Games for {LTL} Fragments}}.
\newblock {\sl \bibinfo{journal}{ACM Transactions on Computational Logic
  (TOCL)}} \bibinfo{volume}{5}(\bibinfo{number}{1}), pp.
  \bibinfo{pages}{1--15}, \doi{10.1145/963927.963928}.

\bibitemdeclare{inproceedings}{BCGH10}
\bibitem{BCGH10}
\bibinfo{author}{R.~Bloem}, \bibinfo{author}{K.~Chatterjee},
  \bibinfo{author}{K.~Greimel}, \bibinfo{author}{T.A. Henzinger} \&
  \bibinfo{author}{B.~Jobstmann} (\bibinfo{year}{2010}):
  \emph{\bibinfo{title}{Robustness in the Presence of Liveness}}.
\newblock In \bibinfo{editor}{T.~Touili}, \bibinfo{editor}{B.~Cook} \&
  \bibinfo{editor}{P.~Jackson}, editors: {\sl \bibinfo{booktitle}{Computer
  Aided Verification (CAV)}}. {\sl \bibinfo{series}{LNCS}}
  \bibinfo{volume}{6174}, \bibinfo{publisher}{Springer},
  \bibinfo{address}{Edinburgh, UK}, pp. \bibinfo{pages}{410--424},
  \doi{10.1007/978-3-642-14295-6\_36}.

\bibitemdeclare{inproceedings}{BCGH10a}
\bibitem{BCGH10a}
\bibinfo{author}{R.~Bloem}, \bibinfo{author}{A.~Cimatti},
  \bibinfo{author}{K.~Greimel}, \bibinfo{author}{G.~Hofferek},
  \bibinfo{author}{R.~Könighofer}, \bibinfo{author}{M.~Roveri},
  \bibinfo{author}{V.~Schuppan} \& \bibinfo{author}{R.~Seeber}
  (\bibinfo{year}{2010}): \emph{\bibinfo{title}{{RATSY} - A New Requirements
  Analysis Tool with Synthesis}}.
\newblock In \bibinfo{editor}{T.~Touili}, \bibinfo{editor}{B.~Cook} \&
  \bibinfo{editor}{P.~Jackson}, editors: {\sl \bibinfo{booktitle}{Computer
  Aided Verification (CAV)}}. {\sl \bibinfo{series}{LNCS}}
  \bibinfo{volume}{6174}, \bibinfo{publisher}{Springer},
  \bibinfo{address}{Edinburgh, UK}, pp. \bibinfo{pages}{425--429},
  \doi{10.1007/978-3-642-14295-6}.

\bibitemdeclare{inproceedings}{BGJP07}
\bibitem{BGJP07}
\bibinfo{author}{R.~Bloem}, \bibinfo{author}{S.~Galler},
  \bibinfo{author}{B.~Jobstmann}, \bibinfo{author}{N.~Piterman},
  \bibinfo{author}{A.~Pnueli} \& \bibinfo{author}{M.~Weiglhofer}
  (\bibinfo{year}{2007}): \emph{\bibinfo{title}{Automatic hardware synthesis
  from specifications: a case study}}.
\newblock In \bibinfo{editor}{R.~Lauwereins} \& \bibinfo{editor}{J.~Madsen},
  editors: {\sl \bibinfo{booktitle}{Design, Automation and Test in Europe
  (DATE)}}. \bibinfo{publisher}{IEEE Computer Society}, \bibinfo{address}{Nice,
  France}, pp. \bibinfo{pages}{1188--1193}.

\bibitemdeclare{article}{BGJP07a}
\bibitem{BGJP07a}
\bibinfo{author}{R.~Bloem}, \bibinfo{author}{S.~Galler},
  \bibinfo{author}{B.~Jobstmann}, \bibinfo{author}{N.~Piterman},
  \bibinfo{author}{A.~Pnueli} \& \bibinfo{author}{M.~Weiglhofer}
  (\bibinfo{year}{2007}): \emph{\bibinfo{title}{Specify, Compile, Run: Hardware
  from {PSL}}}.
\newblock {\sl \bibinfo{journal}{Electronic Notes in Theoretical Computer
  Science (ENTCS)}} \bibinfo{volume}{190}, pp. \bibinfo{pages}{3--16},
  \doi{10.1016/j.entcs.2007.09.004}.

\bibitemdeclare{inproceedings}{BoKu09a}
\bibitem{BoKu09a}
\bibinfo{author}{U.~Boker} \& \bibinfo{author}{O.~Kupferman}
  (\bibinfo{year}{2009}): \emph{\bibinfo{title}{Co-ing {Büchi} Made Tight and
  Useful}}.
\newblock In: {\sl \bibinfo{booktitle}{Logic in Computer Science (LICS)}}.
  \bibinfo{publisher}{IEEE Computer Society}, \bibinfo{address}{Los Angeles,
  California, USA}, pp. \bibinfo{pages}{245--254}, \doi{10.1109/LICS.2009.32}.

\bibitemdeclare{inproceedings}{BCMD90}
\bibitem{BCMD90}
\bibinfo{author}{J.R. Burch}, \bibinfo{author}{E.M. Clarke},
  \bibinfo{author}{K.L. McMillan}, \bibinfo{author}{D.L. Dill} \&
  \bibinfo{author}{L.J. Hwang} (\bibinfo{year}{1990}):
  \emph{\bibinfo{title}{Symbolic Model Checking: $10^{20}$ States and Beyond}}.
\newblock In: {\sl \bibinfo{booktitle}{Logic in Computer Science (LICS)}}.
  \bibinfo{publisher}{IEEE Computer Society}, \bibinfo{address}{Washington, DC,
  USA}, pp. \bibinfo{pages}{1--33}, \doi{10.1109/LICS.1990.113767}.

\bibitemdeclare{inproceedings}{ChMP92}
\bibitem{ChMP92}
\bibinfo{author}{E.Y. Chang}, \bibinfo{author}{Z.~Manna} \&
  \bibinfo{author}{A.~Pnueli} (\bibinfo{year}{1992}):
  \emph{\bibinfo{title}{Characterization of Temporal Property Classes}}.
\newblock In \bibinfo{editor}{W.~Kuich}, editor: {\sl
  \bibinfo{booktitle}{International Colloquium on Automata, Languages and
  Programming (ICALP)}}. {\sl \bibinfo{series}{LNCS}} \bibinfo{volume}{623},
  \bibinfo{publisher}{Springer}, \bibinfo{address}{Vienna, Austria}, pp.
  \bibinfo{pages}{474--486}.

\bibitemdeclare{incollection}{Emer90}
\bibitem{Emer90}
\bibinfo{author}{E.A. Emerson} (\bibinfo{year}{1990}):
  \emph{\bibinfo{title}{Temporal and Modal Logic}}.
\newblock In \bibinfo{editor}{J.~{van Leeuwen}}, editor: {\sl
  \bibinfo{booktitle}{Handbook of Theoretical Computer Science}},
  chapter~\bibinfo{chapter}{16}. \bibinfo{volume}{B: Formal Models and
  Semantics}, \bibinfo{publisher}{Elsevier}, pp. \bibinfo{pages}{995--1072}.

\bibitemdeclare{phdthesis}{Frit05b}
\bibitem{Frit05b}
\bibinfo{author}{C.~Fritz} (\bibinfo{year}{2005}):
  \emph{\bibinfo{title}{Simulation-Based Simplification of omega-Automata}}.
\newblock Ph.D. thesis, \bibinfo{school}{Technischen Fakultät der
  Christian-Albrechts-Universität zu Kiel, Germany}.

\bibitemdeclare{inproceedings}{JGWB07}
\bibitem{JGWB07}
\bibinfo{author}{B.~Jobstmann}, \bibinfo{author}{S.~Galler},
  \bibinfo{author}{M.~Weiglhofer} \& \bibinfo{author}{R.~Bloem}
  (\bibinfo{year}{2007}): \emph{\bibinfo{title}{{Anzu}: A Tool for Property
  Synthesis}}.
\newblock In \bibinfo{editor}{W.~Damm} \& \bibinfo{editor}{H.~Hermanns},
  editors: {\sl \bibinfo{booktitle}{Computer Aided Verification (CAV)}}. {\sl
  \bibinfo{series}{LNCS}} \bibinfo{volume}{4590},
  \bibinfo{publisher}{Springer}, \bibinfo{address}{Berlin, Germany}, pp.
  \bibinfo{pages}{258--262}, \doi{10.1007/978-3-540-73368-3\_29}.

\bibitemdeclare{inproceedings}{KuVa98c}
\bibitem{KuVa98c}
\bibinfo{author}{O.~Kupferman} \& \bibinfo{author}{M.Y. Vardi}
  (\bibinfo{year}{1998}): \emph{\bibinfo{title}{Freedom, Weakness, and
  Determinism: From Linear-Time to Branching-Time}}.
\newblock In: {\sl \bibinfo{booktitle}{Logic in Computer Science (LICS)}}.
  \bibinfo{publisher}{IEEE Computer Society}, \bibinfo{address}{Indianapolis,
  Indiana, USA}, pp. \bibinfo{pages}{81--92}, \doi{10.1109/LICS.1998.705645}.

\bibitemdeclare{inproceedings}{KoHB09}
\bibitem{KoHB09}
\bibinfo{author}{R.~Könighofer}, \bibinfo{author}{G.~Hofferek} \&
  \bibinfo{author}{R.~Bloem} (\bibinfo{year}{2009}):
  \emph{\bibinfo{title}{Debugging formal specifications using simple
  counterstrategies}}.
\newblock In: {\sl \bibinfo{booktitle}{Formal Methods in Computer-Aided Design
  (FMCAD)}}. \bibinfo{publisher}{IEEE Computer Society},
  \bibinfo{address}{Austin, Texas, USA}, pp. \bibinfo{pages}{152--159},
  \doi{10.1109/FMCAD.2009.5351127}.

\bibitemdeclare{inproceedings}{Maid00}
\bibitem{Maid00}
\bibinfo{author}{M.~Maidl} (\bibinfo{year}{2000}): \emph{\bibinfo{title}{The
  Common Fragment of {CTL} and {LTL}}}.
\newblock In: {\sl \bibinfo{booktitle}{Foundations of Computer Science
  (FOCS)}}. pp. \bibinfo{pages}{643--652}.

\bibitemdeclare{inproceedings}{MaPn87c}
\bibitem{MaPn87c}
\bibinfo{author}{Z.~Manna} \& \bibinfo{author}{A.~Pnueli}
  (\bibinfo{year}{1987}): \emph{\bibinfo{title}{A Hierarchy of Temporal
  Properties}}.
\newblock In: {\sl \bibinfo{booktitle}{Principles of Distributed Computing
  (PODC)}}. p. \bibinfo{pages}{205}, \doi{10.1145/41840.41857}.

\bibitemdeclare{inproceedings}{MaPn90}
\bibitem{MaPn90}
\bibinfo{author}{Z.~Manna} \& \bibinfo{author}{A.~Pnueli}
  (\bibinfo{year}{1990}): \emph{\bibinfo{title}{A hierarchy of temporal
  properties}}.
\newblock In: {\sl \bibinfo{booktitle}{Principles of Distributed Computing
  (PODC)}}. \bibinfo{publisher}{ACM}, \bibinfo{address}{Quebec City, Quebec,
  Canada}, pp. \bibinfo{pages}{377--408}.

\bibitemdeclare{article}{MaPn91}
\bibitem{MaPn91}
\bibinfo{author}{Z.~Manna} \& \bibinfo{author}{A.~Pnueli}
  (\bibinfo{year}{1991}): \emph{\bibinfo{title}{Completing the temporal
  picture}}.
\newblock {\sl \bibinfo{journal}{Theoretical Computer Science (TCS)}}
  \bibinfo{volume}{83}(\bibinfo{number}{1}), pp. \bibinfo{pages}{97--130},
  \doi{10.1016/0304-3975(91)90041-Y}.

\bibitemdeclare{article}{MiHa84}
\bibitem{MiHa84}
\bibinfo{author}{S.~Miyano} \& \bibinfo{author}{T.~Hayashi}
  (\bibinfo{year}{1984}): \emph{\bibinfo{title}{Alternating automata on
  $\omega$-words}}.
\newblock {\sl \bibinfo{journal}{Theoretical Computer Science (TCS)}}
  \bibinfo{volume}{32}, pp. \bibinfo{pages}{321--330},
  \doi{10.1016/0304-3975(84)90049-5}.

\bibitemdeclare{inproceedings}{MoSL08}
\bibitem{MoSL08}
\bibinfo{author}{A.~Morgenstern}, \bibinfo{author}{K.~Schneider} \&
  \bibinfo{author}{S.~Lamberti} (\bibinfo{year}{2008}):
  \emph{\bibinfo{title}{Generating Deterministic $\omega$-Automata for most
  {LTL} Formulas by the Breakpoint Construction}}.
\newblock In \bibinfo{editor}{C.~Scholl} \& \bibinfo{editor}{S.~Disch},
  editors: {\sl \bibinfo{booktitle}{Methoden und Beschreibungssprachen zur
  Modellierung und Verifikation von Schaltungen und Systemen (MBMV)}}.
  \bibinfo{publisher}{Shaker}, \bibinfo{address}{Freiburg, Germany}, pp.
  \bibinfo{pages}{119--128}.

\bibitemdeclare{inproceedings}{PiPS06}
\bibitem{PiPS06}
\bibinfo{author}{N.~Piterman}, \bibinfo{author}{A.~Pnueli} \&
  \bibinfo{author}{Y.~{Sa’ar}} (\bibinfo{year}{2006}):
  \emph{\bibinfo{title}{Synthesis of Reactive(1) Designs}}.
\newblock In \bibinfo{editor}{E.A. Emerson} \& \bibinfo{editor}{K.S. Namjoshi},
  editors: {\sl \bibinfo{booktitle}{Verification, Model Checking, and Abstract
  Interpretation (VMCAI)}}. {\sl \bibinfo{series}{LNCS}}
  \bibinfo{volume}{3855}, \bibinfo{publisher}{Springer},
  \bibinfo{address}{Charleston, South Carolina, USA}, pp.
  \bibinfo{pages}{364--380}, \doi{10.1007/11609773\_24}.

\bibitemdeclare{inproceedings}{Pnue77a}
\bibitem{Pnue77a}
\bibinfo{author}{A.~Pnueli} (\bibinfo{year}{1977}): \emph{\bibinfo{title}{The
  Temporal Logic of Programs}}.
\newblock In: {\sl \bibinfo{booktitle}{Foundations of Computer Science
  (FOCS)}}. \bibinfo{publisher}{IEEE Computer Society},
  \bibinfo{address}{Providence, Rhode Island, USA}, pp.
  \bibinfo{pages}{46--57}, \doi{10.1109/SFCS.1977.32}.

\bibitemdeclare{inproceedings}{Schn01b}
\bibitem{Schn01b}
\bibinfo{author}{K.~Schneider} (\bibinfo{year}{2001}):
  \emph{\bibinfo{title}{Improving Automata Generation for Linear Temporal Logic
  by Considering the Automata Hierarchy}}.
\newblock In \bibinfo{editor}{R.~Nieuwenhuis} \& \bibinfo{editor}{A.~Voronkov},
  editors: {\sl \bibinfo{booktitle}{Logic for Programming, Artificial
  Intelligence, and Reasoning (LPAR)}}. {\sl \bibinfo{series}{LNAI}}
  \bibinfo{volume}{2250}, \bibinfo{publisher}{Springer},
  \bibinfo{address}{Havana, Cuba}, pp. \bibinfo{pages}{39--54},
  \doi{10.1007/3-540-45653-8\_3}.

\bibitemdeclare{book}{Schn03}
\bibitem{Schn03}
\bibinfo{author}{K.~Schneider} (\bibinfo{year}{2003}):
  \emph{\bibinfo{title}{Verification of Reactive Systems - Formal Methods and
  Algorithms}}.
\newblock \bibinfo{series}{Texts in Theoretical Computer Science (EATCS
  Series)}, \bibinfo{publisher}{Springer}.

\bibitemdeclare{techreport}{Schn09}
\bibitem{Schn09}
\bibinfo{author}{K.~Schneider} (\bibinfo{year}{2009}):
  \emph{\bibinfo{title}{The Synchronous Programming Language {Quartz}}}.
\newblock \bibinfo{type}{Internal Report} \bibinfo{number}{375},
  \bibinfo{institution}{Department of Computer Science, University of
  Kaiserslautern}, \bibinfo{address}{Kaiserslautern, Germany}.

\bibitemdeclare{incollection}{Thom90a}
\bibitem{Thom90a}
\bibinfo{author}{W.~Thomas} (\bibinfo{year}{1990}):
  \emph{\bibinfo{title}{Automata on Infinite Objects}}.
\newblock In \bibinfo{editor}{J.~{van Leeuwen}}, editor: {\sl
  \bibinfo{booktitle}{Handbook of Theoretical Computer Science}},
  chapter~\bibinfo{chapter}{4}. \bibinfo{volume}{B: Formal Models and
  Semantics}, \bibinfo{publisher}{Elsevier}, pp. \bibinfo{pages}{133--191}.

\bibitemdeclare{article}{Wagn79}
\bibitem{Wagn79}
\bibinfo{author}{K.~Wagner} (\bibinfo{year}{1979}): \emph{\bibinfo{title}{On
  $\omega$-regular sets}}.
\newblock {\sl \bibinfo{journal}{Information and Control}}
  \bibinfo{volume}{43}(\bibinfo{number}{2}), pp. \bibinfo{pages}{123--177}.

\bibitemdeclare{inproceedings}{WaHT03}
\bibitem{WaHT03}
\bibinfo{author}{N.~Wallmeier}, \bibinfo{author}{P.~Hütten} \&
  \bibinfo{author}{W.~Thomas} (\bibinfo{year}{2003}):
  \emph{\bibinfo{title}{Symbolic Synthesis of Finite-State Controllers for
  Request-Response Specifications}}.
\newblock In \bibinfo{editor}{O.H. Ibarra} \& \bibinfo{editor}{Z.~Dang},
  editors: {\sl \bibinfo{booktitle}{Conference on Implementation and
  Application of Automata (CIAA)}}. {\sl \bibinfo{series}{LNCS}}
  \bibinfo{volume}{2759}, \bibinfo{publisher}{Springer},
  \bibinfo{address}{Santa Barbara, California, USA}, pp.
  \bibinfo{pages}{11--22}, \doi{10.1007/3-540-45089-0\_3}.

\bibitemdeclare{inproceedings}{WoTM10}
\bibitem{WoTM10}
\bibinfo{author}{T.~Wongpiromsarn}, \bibinfo{author}{U.~Topcu} \&
  \bibinfo{author}{R.M. Murray} (\bibinfo{year}{2010}):
  \emph{\bibinfo{title}{Receding horizon control for temporal logic
  specifications}}.
\newblock In \bibinfo{editor}{K.H. Johansson} \& \bibinfo{editor}{W.~Yi},
  editors: {\sl \bibinfo{booktitle}{Hybrid Systems: Computation and Control
  (HSCC)}}. \bibinfo{publisher}{ACM}, \bibinfo{address}{Stockholm, Sweden}, pp.
  \bibinfo{pages}{101--110}, \doi{10.1145/1755952.1755968}.

\end{thebibliography}
\end{document}